\documentclass[twocolumn,showpacs,superscriptaddress,pra,appendix]{revtex4}

\usepackage{mathrsfs}
\usepackage{amsmath}
\usepackage{amssymb}
\usepackage{graphicx}
\usepackage{dcolumn}
\usepackage{bm,braket}
\usepackage{subfigure}
\usepackage{color}
\usepackage{ifpdf}
\usepackage{epstopdf}
\usepackage{CJK}
\usepackage{float}

\begin{document}

\title{Exact results for polaron and molecule   in one-dimensional spin-1/2 Fermi gas}
\begin{CJK*}{UTF8}{gbsn}

\author {Runxin Mao(毛润欣)}
\affiliation{International Center for Quantum Materials, School of Physics, Peking University, Beijing 100871, China}

\author{X. W. Guan(管习文)}
\affiliation{State Key Laboratory of Magnetic Resonance and Atomic and Molecular Physics, Wuhan Institute of Physics and Mathematics, Chinese Academy of Sciences, Wuhan 430071, China}
\affiliation{Department of Theoretical Physics, Research School of Physics and Engineering,
Australian National University, Canberra ACT 0200, Australia}

\author{Biao Wu(吴飙)}
     \email{wubiao@pku.edu.cn}
      \affiliation{International Center for Quantum Materials, School of Physics, Peking University, Beijing 100871, China}
      \affiliation{Collaborative Innovation Center of Quantum Matter, Beijing 100871, China}
      \affiliation{Wilczek Quantum Center, College of Science, Zhejiang University of Technology,  Hangzhou 310014, China}
       \affiliation{Synergetic Innovation Center for Quantum Effects and Applications, Hunan Normal University, Changsha 410081, China}

\begin{abstract}
Using exact Bethe ansatz (BA) solutions, we show that a spin-down fermion  immersed into
a fully polarized spin-up Fermi sea with a weak attraction is dressed by the surrounding spin-up fermions to form the  one-dimensional analog  of a  polaron. As the attraction becomes strong, the spin-down fermion binds with one spin-up fermion  to form a tightly bound molecule.  Throughout the whole interaction regime, a  crossover from the polaron to a molecule state is fully demonstrated through   exact results of the excitation spectrum, the  effective mass, binding energy and kinetic energy.  Furthermore,  a clear distinction between the polaron and molecule is conceived by the probability distribution, single particle reduced density matrix   and density-density correlations, which are calculated directly from  the Bethe ansatz wave function. Such a polaron-molecule crossover presents  a universal nature of an impurity immersed into a fermionic medium with an attraction in one dimension.
\end{abstract}


\maketitle

\end{CJK*}


\section{Introduction}
Advance in trapping and manipulating cold fermionic atoms have provided an experimental
realization of various many-body phenomena~\cite{Lewenstein2007,Giorgini2008,Cazalilla:2011,Guan:2013a}. In particular, the recent observation of Fermi polarons in a three-dimensional (3D) tunable
Fermi liquid of ultracold atoms~\cite{Schirotzek2009,Navon2010,Kohstall:2012} provides
insightful understanding of quasiparticle physics in many-body systems~\cite{Massignan:2014}.
The Fermi polaron is a dressed spin-down impurity fermion by the surrounding
scattered fermions in a spin-up Fermi sea. With increasing attraction, the single spin-down
fermion  undergoes a possibly polaron-molecule transition in the fermionic medium in 3D. The study of quasiparticle physics and the dynamics of polarons and molecules in fermionic  medium
has received much theoretical and experimental attention~\cite{Schirotzek2009,Navon2010,NAscimbene2009,Combescot2009,Bruum2010,Mathy2013,Levinsen:2012,Yi:2015,Schmidt:2016}. In this context, all studies concerning the first order nature of the polaron-molecule transition in a 3D fermionic medium~\cite{NAscimbene2009,Combescot2009,Bruum2010} involve variational ansatz with some approximations that is ultimately not justified in low
dimensions~\cite{Parish2011,Giraud:2009}. It is therefore highly desirable to have some rigorous results of such quasiparticle physics in different mediums.

\begin{figure}
\includegraphics[width=0.95\linewidth]{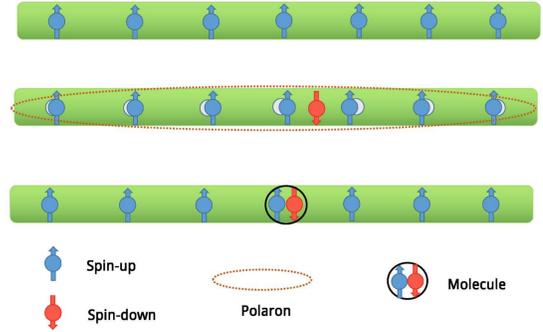}
\caption{ (color online) Schematic configuration of polaron-molecule crossover of a single attractive
impurity in the 1D  Fermi gas.
In weak attractive limit, the single impurity (red), a spin-down fermion,  dressed
by the surrounding scattered spin-up fermions (blue) from the medium behaves like  a polaron
(dashed oval) with an effective mass \(m^*\)$\approx $m. For strong attraction, it binds
with one spin-up fermion from the Fermi sea to a tightly bound molecule (black circle)
of two-atom with  an effective mass \(m^*\) $ \approx $2m.
}
\label{schemactic}
\end{figure}

For  1D systems, there in general does not exhibit
true quasiparticles due to strong collective nature at low temperatures.
However, such a collective behavior  does not  rule out the existence of
polarons for very few impurities immersed into the bosonic or fermionic mediums.
In contrast to the Fermi liquid theory for the study of the Fermi polaron in 2D/3D,
the exact Bethe ansatz solutions are more capable of capturing microscopic origin
of polarons and molecules~\cite{McGuire:1965,Guan-FP,Astrakharchik:2013,Dehkharghani:2015,Gharashi:2015,Loft:2016,Doggen:2014}.  In this regard, the 1D spin-1/2 delta-function
interacting Fermi gas~\cite{Yang1967,Gaudin1967}  is ideal for the study of quantum
impurity problem~\cite{Guan:2013a,McGuire:1965,Guan-FP}. The fundamental
physics of this model with arbitrary spin population imbalance is determined
by a set of transcendental equations which were found by  Yang  using
Bethe ansatz (BA) hypothesis in 1967~\cite{Yang1967}. This model  shows many
interesting physical properties~\cite{Guan:2013a}.  Here we show that  the exact BA solutions
can be used to study the different properties  of polarons and molecules in the 1D Fermi gases.

In this paper, using exact  Bethe Ansatz (BA) solution, we rigorously
study Fermi polaron and molecule states in a 1D fermionic medium.
For weak attraction, the single spin-down impurity gets dressed by
the surrounding spin-up fermions to form a polaron-like quasiparticle.
However, as the attractive interaction grows, the spin-down fermion binds
only one spin-up fermion from the medium to gradually form a tightly bound molecule.
See a cartoon picture  shown in Fig.~\ref{schemactic}. In comparison with the previous study~\cite{Guan-FP},
here  we analytically and numerically calculate the polaron  energy, the binding energy,
the effective mass for this impurity problem. Moreover, we  obtain an explicit form of the BA
wave function of the single spin-down immersed in the fully-polarized Fermi sea.
Using this  exact  wave function we further calculate  the distributions and
the correlations of the polaron and molecule which are the major quantities for experimental
measurements on these states.   Our result provides a microscopic origin of the polarons
and molecules resulting in from quantum impurities.

The   paper is organized  as follows: in Secton II, we derive explicit form of
the Bethe wave function. In Section III, we analytically and numerically calculate
the excitation spectrum with a precise determination of the effective mass and
binding energy. In section IV and V we calculate the probability distribution
functions and correlation functions.  We conclude in Section VI.


\section{Model and Beth wave function}
We study the  one dimensional spin-1/2 Fermi gas called Yang-Gaudin model \cite{Yang1967,Gaudin1967}, where the fermions interact with each other
via  the $\delta$-function potential.  Due to the symmetry of the wave function,
the interaction only occurs  between two fermions with different spins. The Hamiltonian of the  system thus has
the following form~\cite{Yang1967,Gaudin1967}
\begin{eqnarray}
H&=&\sum_{\sigma=\downarrow,\uparrow}\int\phi_{\sigma}^{+}(x)(-\frac{\hbar^{2}}{2m}\frac{d^{2}}{dx^{2}})\phi_{\sigma}(x)\nonumber\\
&&+g_{1D}\int\phi_{\downarrow}^{+}(x)\phi_{\uparrow}^{+}(x)\phi_{\downarrow}(x)\phi_{\uparrow}(x)\,,
\label{ham}
\end{eqnarray}
where $m$ is the atomic mass, $g_{1D}$ characterizes the strength of the $\delta$-function interaction, and
the field operators  $\phi_{\uparrow}$ and $\phi_{\downarrow}$  describe the fermionic atoms in the states $\ket{\uparrow}$ and $\ket{\downarrow}$, respectively.  In this work we focus
on the case where a single spin-down  fermion resides in the sea of $N-1$ spin-up fermions.

The system described by the Hamiltonian Eq.(\ref{ham}) is a prototypical integrable model, which  has been experimentally realized with ultracold atoms  trapped in 1D geometry~\cite{Liao:2010,Wenz:2013}. In such 1D  experiments, the wave guide atoms are tightly confined in two transverse directions and weakly confined in the axial direction.
Consequently,  the trapped atoms can be effectively described by the Hamiltonian (\ref{ham}) within the local density approximation ~\cite{Liao:2010,Wenz:2013,Pagano:2014} or by  exact strong coupling ansatz wave functions of the trapped gas~\cite{Loft:2016,Seyed:2013,Deuretzbacher2014,Volosniev2015}.
In these experiments, the coupling constant $g_{1D}$ can be written as $g_{1D}=\hbar^{2}c/m$ with $c=-2/a_{1D}$, where \(a_{1D}\) is the effective 1D scattering length.  According to Ref.~\cite{Olshanii1998},  \(a_{1D}\) is related to the 3D scattering length
\(a_{3D}\)  as \(a_{1D}=-a_{+}^2/a_{3D}+Aa_{+}\), where  \(a_{+}=\sqrt{\hbar/(m\omega_{+})}\) is the transverse oscillator length, and A$\approx$1.0326 is a  constant. For repulsive fermions, c\(>\)0 and  for attractive fermions, c\(<\)0.  The Bethe ansatz solutions of the model provide a precise understanding of  many-body phenomena and few-body physics, see review \cite{Guan:2013a}.   
In this paper we focus on the model (\ref{ham})  with attractive interaction and periodic boundary conditions.

The Bethe wave function of the model Eq.(\ref{ham}) is very complicated.
{ With the help of Takahashi's Bethe ansatz wave function}  \cite{Takahash1999}, we first simplify
the Bethe wave function of an energy eigenstate for the case with the $N$th
fermion being spin-down, i.e.
 \begin{equation}
f_{\downarrow_N}(x_{1},x_{2},...,x_{N})=\sum_{l=1}^{N}\left|\begin{array}{ccc}
a_{11} & ... & a_{1,N-1}\\
. & . & .\\
. & a_{ij} & .\\
. & . & .\\
a_{N-1\text{,}1} & ... & a_{N-1,N-1}\end{array}\right|e^{ik_{l}x_{N}}\,,
\label{wave function}
\end{equation}
where
\begin{equation}
a_{ij}=[k_{l+j}-\lambda+ic'\mathrm{sign}(x_{N}-x_{i})]e^{ik_{l+j}x_{i}}
\end{equation}
if $l+j\leq N$; for other cases,
\begin{equation}
a_{ij}=[k_{l+j-n}-\lambda+ic'\mathrm{sign}(x_{N}-x_{i})]e^{ik_{l+j-N}x_{i}}\,.
\end{equation}
 The wave function for the four-body case is given  explicitly
in Appendix A. In the above wave function,  we denote $c'=c/2$, and $\lambda$ is  the
spin rapidity parameter and $\{k_j\}$ with $j=1,2,\ldots, N$  are the quasi-momenta
of fermions. They can be determined by  the Bethe Ansatz
equations~\cite{Yang1967} (also see below).

The total wave function of our system should be anti-symmetric under
the permutation of any two fermions.  To construct this total wave function,
we define
\begin{equation}
f_{\downarrow_j}(x_{1},...,x_{j},...,x_{N})=-f_{\downarrow_N}(x_{1},...,x_{N},...,x_{j})\,,
\end{equation}
where the subscript $\downarrow_j$ means that the $j$th fermion is spin-down while
the rest of the fermions are spin-up.
As a result,  the total wave function takes the following form,
\begin{equation}
f_{tot}=\frac{1}{\sqrt{G}}\sum_{j=1}^N f_{\downarrow_j}\ket{\downarrow_j}\,,
\label{total}
\end{equation}
and $G$ is the normalization constant and $\ket{\downarrow_j}$ denotes a spin state
with the $j$th spin down and all other spins up.

For solving our problem, we write down  the following
the BA equations~\cite{Yang1967}
\begin{eqnarray}
&&\frac{k_j-\lambda+ic'}{k_j-\lambda-ic'}=\exp(ik_jL)\label{ba01}\,,\\
&&\prod_{j=1}^N\frac{k_j-\lambda+ic'}{k_j-\lambda-ic'}=1\,.
\label{ba02}
\end{eqnarray}
The corresponding eigen-energy  is given by
\begin{equation}
 E=\sum_{j=1}^N \frac{\hbar^2k_j^2}{2m}\,.
\end{equation}
This problem of $N-1$ fermions of the up spin and one fermion of the opposite spin
was studied by McGuire  in 1965 and 1966~\cite{McGuire1965,McGuire1966}, who
calculated only the energy shift caused by the extra spin-down fermion. However, the key
feature of this impurity problem is the collective behavior of polaron and molecule,
which lacks a comprehensive understanding.  In the rest of the paper (except Section IV),
for convenience and without loss of generality, we consider only the cases where  $N$ is even.

\section{Polaron and Molecule}
When a single attractive impurity is immersed in the fully-polarized Fermi sea,
an intuitive picture immediately arises as illustrated schematically in Fig. \ref{schemactic}.
The single impurity will be addressed by  a cloud of fermions due to attraction.
When the impurity moves, it will drag these fermions along.
Effectively, it can be regarded as a new particle with a different mass moving freely.
This is the well known polaron. When the attraction becomes very strong,
the impurity can pair with one fermion to form a molecule.
This nature is reflected by Haldane generalized
exclusion statistics~{\cite{Haldane:1991,Wu:1994,Ha:1994,Murthy:1994,Nayak1994}}, i.e.
the statistical interaction and dynamical interaction are transmutable in 1D.
For a weak attraction,we see that the quasimomentum of a spin-down fermion essentially
depends on that of all spin-up fermions. This gives  a statistical nature of polaron,
namely mutual statistics~\cite{Wu:1994,Batchelor:2006}. Whereas for a strong attraction,
the spin-down fermion tightly bounds with a spin-up fermion from the fully-polarized
Fermi sea. In this circumstance, the statistics of the bond pair reveals a non-mutual statistics.

To rigorously establish such a  transition,  we examine the energy shift caused by the impurity
and see how it changes with the interaction strength.
When there is no interaction, the single spin-down fermion can share a momentum
with a spin-up fermion. This implies mathematically that there exists a pair of $k$'s, say,
$k_{N-1}$ and $k_{N}$,  such that $k_{N-1}=k_{N}=p$. When the attractive interaction
between spin-up fermion and spin-down fermion is turned on, these two momenta can
become a pair of complex conjugates, \(k_{N-1,N}=p\pm\ i\beta\), indicating
the formation of a polaron or molecule.

\subsection{Weak coupling}
We first consider the weak coupling limit,  \(L|c|\ll1\).  In this limit, it is straightforward to find
that $p\approx \lambda$ and $\beta\approx \sqrt{|c|/L}$. With these two facts~\cite{Guan-FP},   we
can readily obtain from Eq.(\ref{ba01}) and Eq.(\ref{ba02})
\begin{eqnarray}
p&\approx&\frac{2n_p\pi}{L}-\frac{1}{2}\sum_{j=1}^{N-2}\frac{|c|}{2(n_p-n_j)\pi}\\
k_j&\approx&\frac{2n_j\pi}{L}-\frac{|c|}{2(n_j-n_p)\pi}\,,~~~~(j=1,...,N-2)\,,
\end{eqnarray}
where $n_j$'s and $n_p$ are integers and $n_j\neq n_p$.
According to the Fermi statistics,  for the lowest energy state, we have \(n_j=\pm1,\pm2...(N-2)/2\)
(for details see Appendix B).  From the above equations, we can calculate the energy
of the system with a single spin-down fermion
\begin{eqnarray}
E &=&\frac{\hbar^{2}}{2m}\Big( -2\beta^{2}+2p^{2}+\sum_{j=1}^{N-2}k_{j}^{2}\Big)\nonumber\\
&\approx&-\frac{\hbar^{2}}{2m}\frac{2(N-1)|c|}{L}+
\frac{\hbar^{2}q^{2}}{2m}+E_{0},
\label{polar}
\end{eqnarray}
where $q=2n_{p}\pi/L$ and $E_0$ is the ground state energy of the system when
there is no interaction, that is, $c=0$.  For \(n_p=0\), we recover the ground
state energy of the gas with $N-1$ spin-up fermions and one spin-down fermion
given by McGuire~\cite{McGuire1965,McGuire1966}.

The energy expression (\ref{polar}) naturally gives the dispersion of a
polaron-like quasiparticle moving slowly in the fully polarized Fermi sea.
{It  provides a rich  insight into such collective nature of polaron: 1)  the first term is the mean  binding energy, showing that the spin-down
fermion experiences a mean field attraction from the fully polarized Fermi sea;
2) the second term is the kinetic energy of a single classical particle with momentum
$\hbar q$ and mass $m$.}  When $c=0$, this is just the kinetic energy for the newly
added spin-down fermion. When $c\neq 0$ but small,  this kinetic energy does not
change its form even though the spin-down fermion is interacting with
spin-up fermions and has lost its individual character.  {In this regard, one can view
the  spin-down fermion addressed by the surround spin-up fermions as a polaron. In one dimension, such a polaronic behaviour  is a typical elementary excitation with infinite lifetime due to the reshuffle  of the eigenstates in excitations}.  When $c$ becomes
bigger, the effective mass of the polaron will be different from the bare
mass $m$ (the leading order contribution is  proportional to $c^2$), but
overall picture remains the same: a Fermi sea, a binding energy $E_{b}=-\frac{\hbar^{2}}{2m}\frac{2(N-1)|c|}{L}$, and a classical particle with an effective mass $m^*$.

The above results are valid up to the first order of $cL$. The results to the second order
of $cL$ can be found in the Appendix B. There is a correction to the binding energy. {However, it is very hard to calculate  the effective mass of the polaron for the order over the first order of  $cL$. }

\subsection{Strong coupling}
We now consider the strong coupling limit,  \(|c|L\gg1\).
Using $(|c|L)^{-1}$ as the perturbation parameter \cite{Guan-FP},
we find $p$ and the $N-2$ real momenta
\begin{eqnarray}
p&\approx& \frac{n_{p}\pi}{L}-\sum_{j=1}^{N-2}\frac{2n_{j}\pi}{|c|L}+\frac{2(N-2)n_{p}\pi}{|c|L}\\
k_{j}&\approx &\frac{n_{j}\pi}{L}+\frac{4n_{j}\pi-4n_{p}\pi}{|c|L}
\end{eqnarray}
with $n_{j}=\pm1,\pm3,....\pm($$N-3)$ and $n_p$ is an arbitrary integer.  In addition,
we find  $\beta\approx|c|/2$. Therefore, the total energy is
\begin{eqnarray}
E&=&\frac{\hbar^{2}}{2m}\Big(-2\beta^{2}+2p^{2}+\sum_{j=1}^{N-2}k_{j}^{2}\Big)\nonumber\\
&=&-\frac{\hbar^{2}}{2m}\Big[\frac{c^{2}}{2}+(1+\frac{8}{|c|})\frac{(2N-1)(N-2)\pi^{2}}{L^2}\Big]\nonumber\\
&&+\Big[\frac{1}{2}+\frac{2(N-2)}{|c|}\Big]\frac{\hbar^{2}q^{2}}{2m}+(1+\frac{8}{|c|})E_{0}\,.\label{molecule}
\label{moleculeenergy}
\end{eqnarray}
Similar to the case of weak coupling,  this energy has three terms: the binding energy,
the kinetic energy, and the Fermi sea energy. However, all of them are different, and
even the Fermi sea energy is slightly modified by the strong interaction. In particular,
the kinetic energy can be regarded as a quasi-particle with effective mass
\begin{equation}
m^{*}\approx2m\left(1-\frac{4(N-2)}{L|c|}\right)\,, \label{effective-M-S}
\end{equation}
which is almost twice the bare mass $m$.  After a subtraction of the ground state energy
and chemical potential within Eq. (\ref{molecule}), we obtain the binding energy
of the molecule state
\begin{equation}
E_{b}=\frac{\hbar^{2}}{2m}\left\{- \frac{c^2}{2}+\frac{8\pi^2 }{3|\gamma|}\right\}\,, \label{Eb-strong}
\end{equation}
where $\gamma=c/n$  is a dimensionless interaction strength

\begin{figure}
\includegraphics[width=0.95\linewidth]{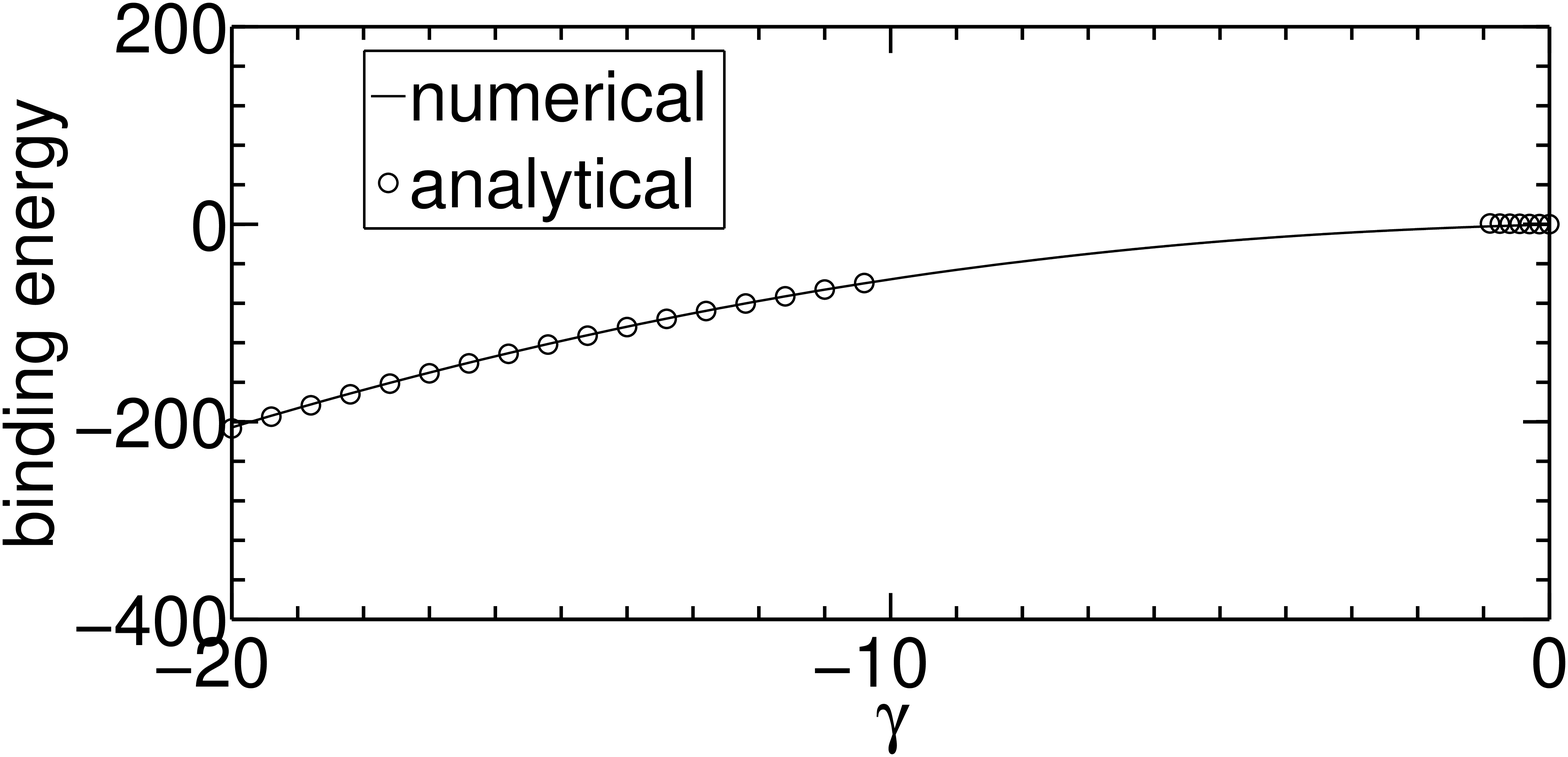}
\caption{Binding energy $E_{b}$. The solid line is the numerical result and
the  circles stand the result from the  analytical expression  (\ref{Eb-strong}) { and (\ref{bindingE-w}) for strong and weak attractions, respectively}.
The unit of energy is  $\frac{\hbar^{2}n^{2}}{2m}$ { and $\gamma$ is the dimensionless interaction strength.}
$N=10$, $L=1$.
}
\label{binding energy}
\end{figure}

\begin{figure}
\includegraphics[width=0.95\linewidth]{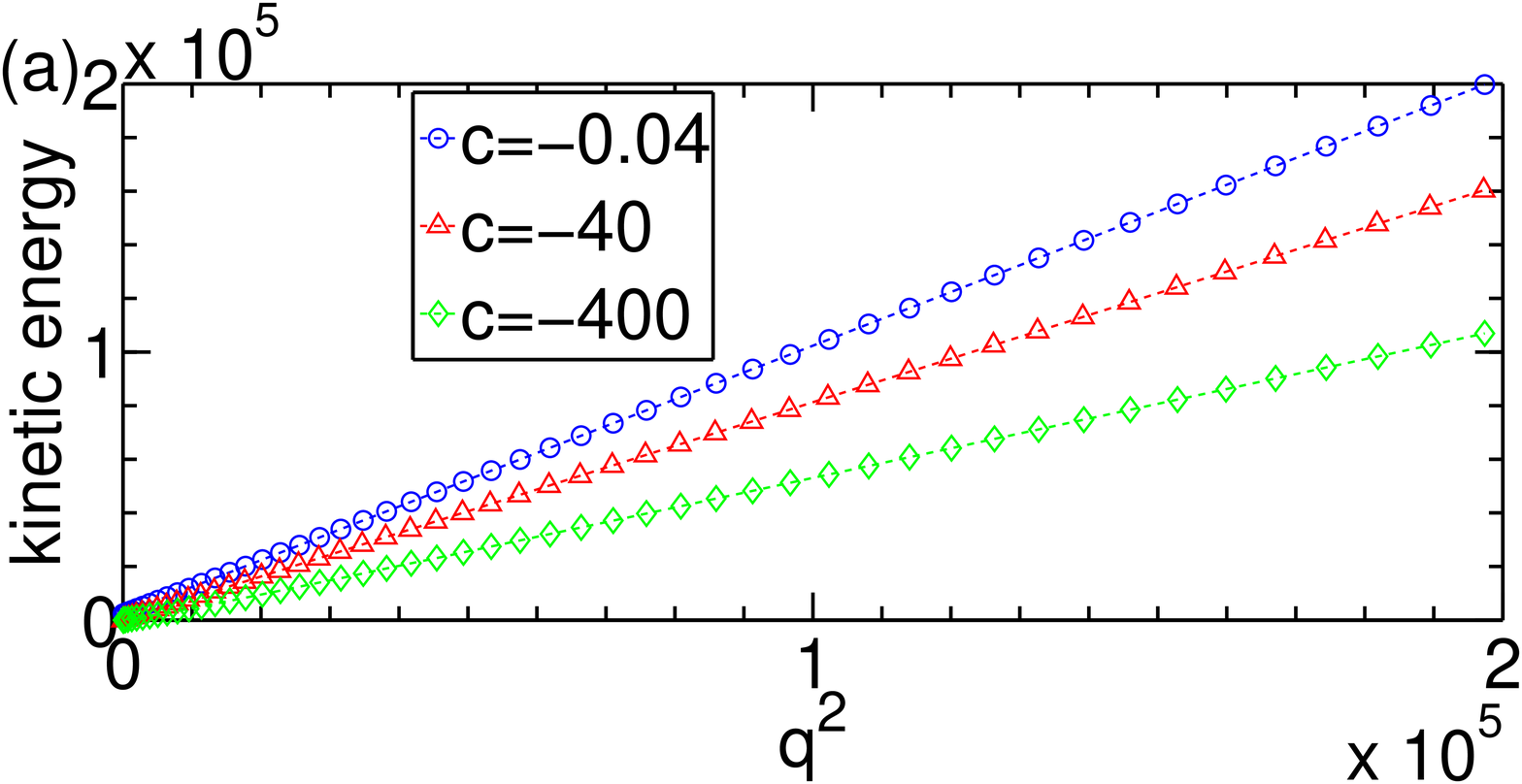}
\includegraphics[width=0.95\linewidth]{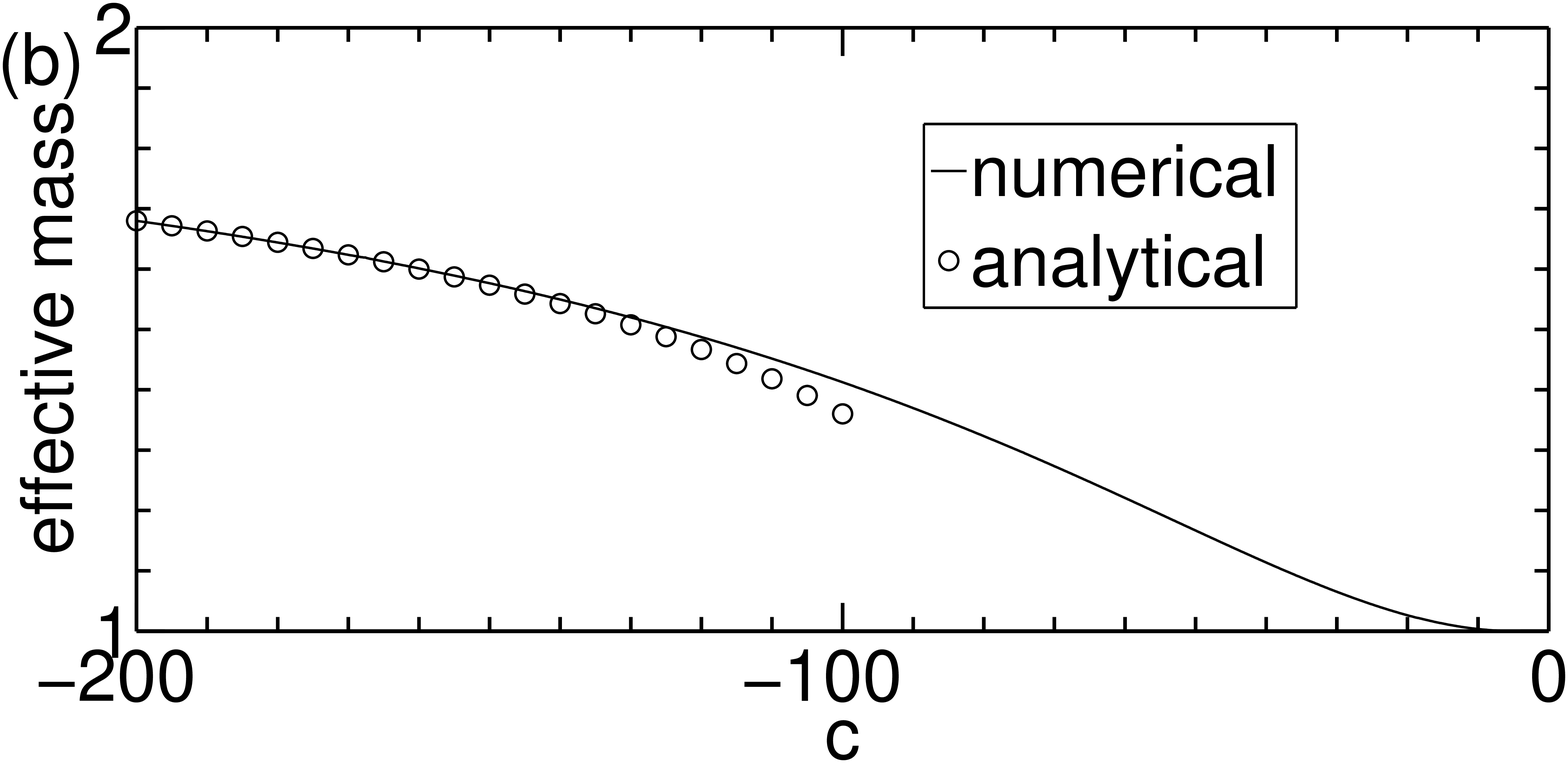}
\caption{(color online) (a) The kinetic energy as a function of the square of the momentum $q$
for different values of interaction strength. The unit of energy is  $\hbar^2/(2mL^2)$ { and the unit of $q^2$ is $L^{-2}$}.
(b) Effective mass $m^{*}/m$  as a function of interaction strength $c$.  The line is the
numerical result and the circles denote the analytical result obtained from (\ref{effective-M-S})
for strong coupling regime.  For weak coupling regime, the leading order contribution
to the effective mass is identified as an order of $c^2$. {The  mass ratio $m^{*}/m$  has no unit and the unit of $c$ is $L^{-1}$}.  Here $N=10$ and $L=1$. }
\label{effective mass}
\end{figure}

\subsection{General case}
The above two limiting cases show that the total energy of the system can be
divided into three parts,
\begin{eqnarray}
E=E_0+E_{b}+\frac{\hbar^{2}q^{2}}{2m^{*}}\,.
\label{energystate}
\end{eqnarray}
$E_0$ is  the Fermi sea energy,  the ground state energy of $N$ free fermions; $E_b$ is
the binding energy;  the last term is  the kinetic energy of the quasi-particle, i.e. dispersion of  a polaron.
In the weak coupling limit, the binding energy is
\begin{eqnarray}
E_{b}\approx-\frac{\hbar^{2}}{2m}\frac{2|c|}{L}(N-1)=2\gamma \frac{\hbar^{2}n^{2}}{2m}\,,
\label{bindingE-w}
\end{eqnarray}
where $n=N/L$ is the density of particle.
In the strong coupling limit, the binding energy is given by (\ref{Eb-strong}).

{A crossover from a polaron to the molecule   is seen for an intermediate interaction strength, i.e.  the total energy of this system can be divided
into  three parts as indicated in (\ref{energystate}). } However, in general, the BA equations in Eq.(\ref{ba01},\ref{ba02})
can not be solved analytically. We have to resort to numerical method.  To compute
the binding energy $E_b$, we compute the ground state of the system and then
subtract out $E_0$. The numerical results with $N=10$, $L=1$ are shown in
Fig. \ref{binding energy}.  {We observe
 that our numerical results  agree well with  analytical expressions (\ref{Eb-strong}) and (\ref{bindingE-w}) in  the weak and strong coupling limits. }

For the kinetic energy, we compute the system energy for a given momentum $q$ and then
subtract out $E_0$ and $E_b$. The results are plotted in Fig.~\ref{effective mass} (a), where
the kinetic energy is seen to grow as a power of $q^2$. By extracting the slope,
we can compute the effective mass $m^*$, which is  shown in Fig.~\ref{effective mass} (b).
{ In  the weak coupling limit, we indeed see that $m^*$ approaches to the bare mass $m$.
In  the strong coupling limit, the numerical result of $m^*$ agrees well with our analytical result (\ref{effective-M-S}), } i.e.
$m^{*}\approx2m(1-\frac{4}{|\gamma|})$. These results fully confirm the
polaron behavior in the problem of such a spin-down fermion
immersed into a fully-polarized Fermi sea in 1D.

\section{Probability distribution function of three fermions}
 \begin{figure}
\includegraphics[width=0.95\linewidth]{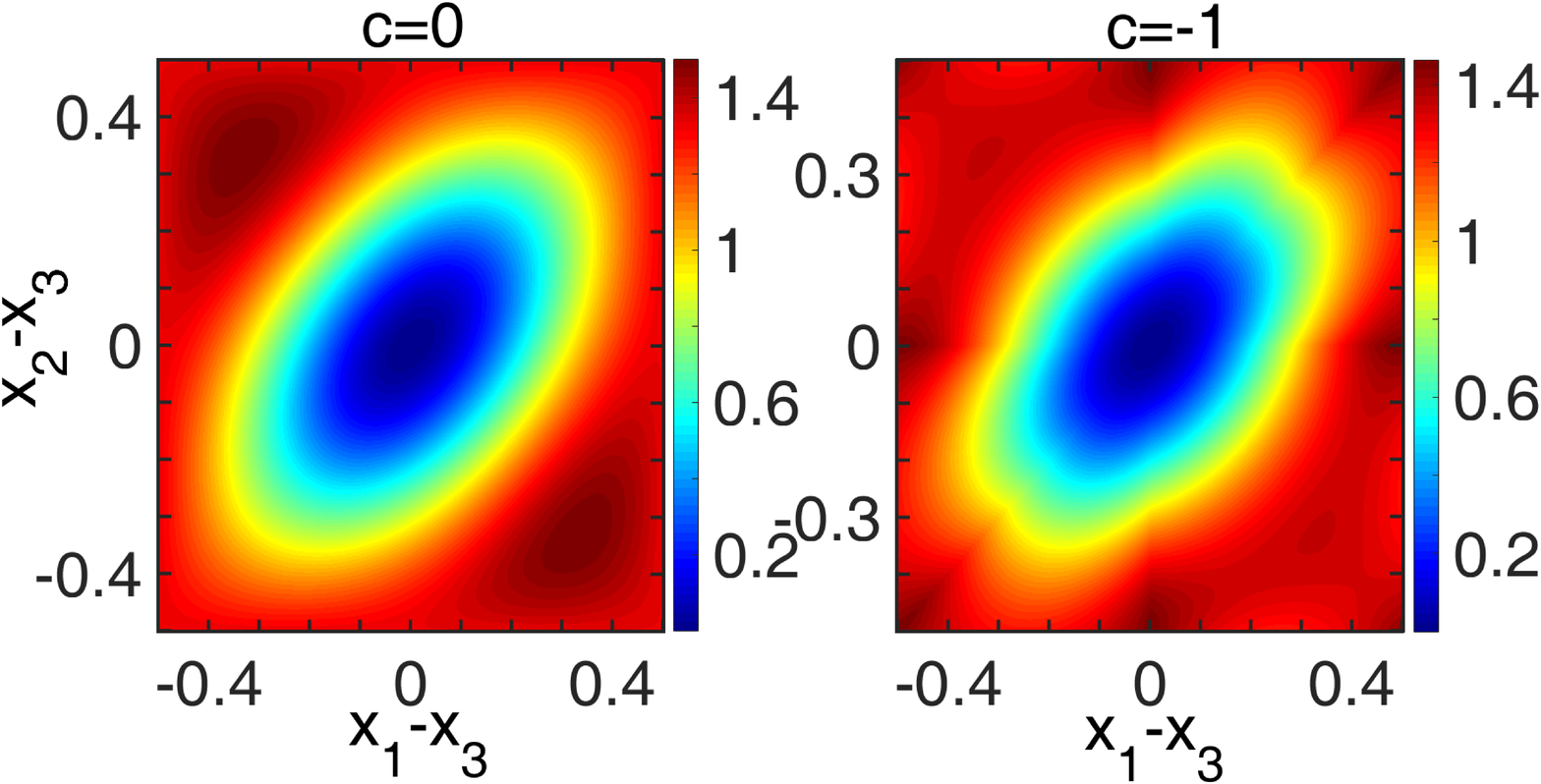}
\includegraphics[width=0.95\linewidth]{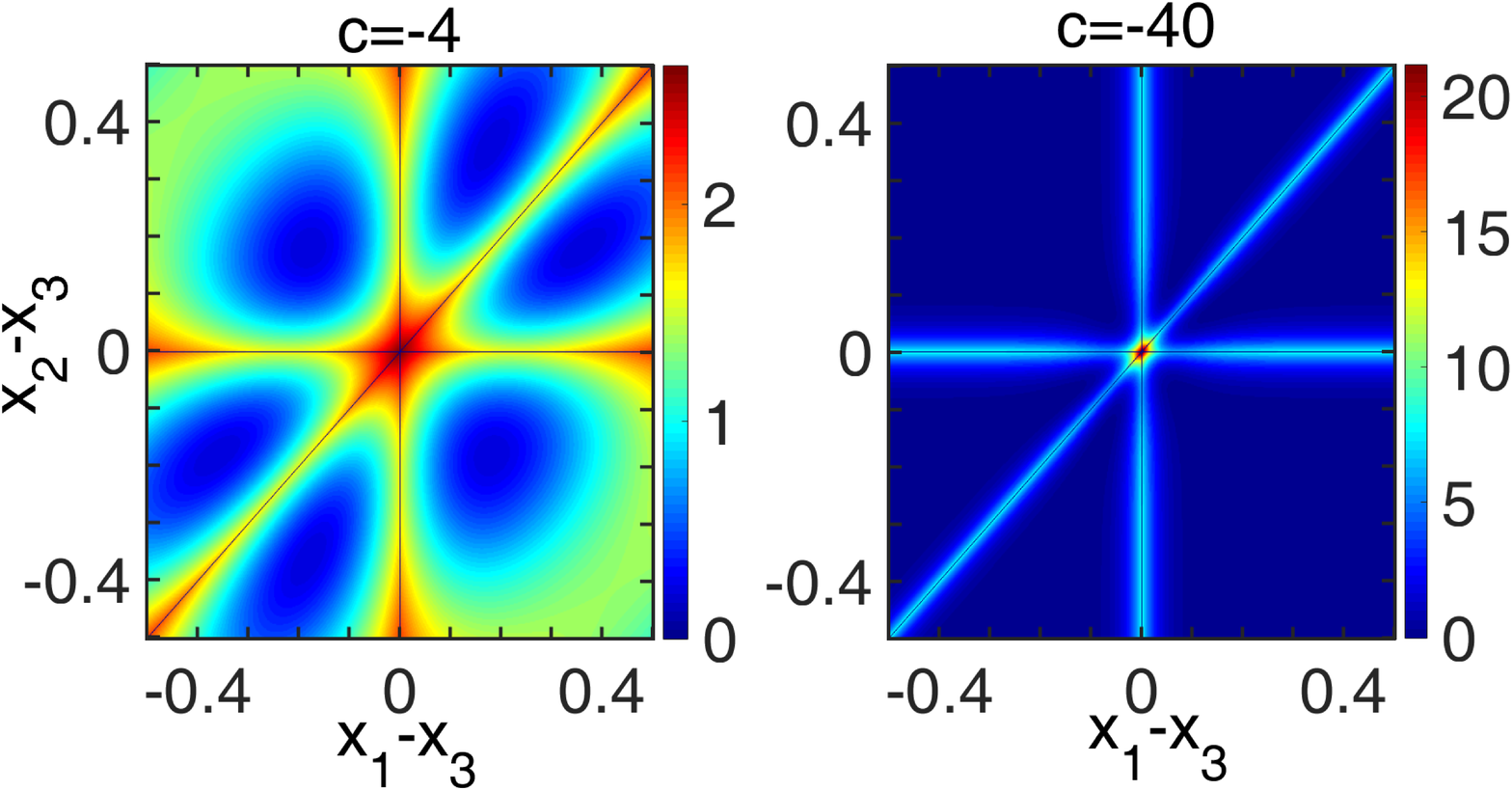}
\caption{(color online) Probability distribution of the ground state in the case of
three fermions for $ c=0, -1, -4, -40$.
When $c=0$, the ground state corresponds to $\{k_1,k_2,k_3\}=\{0,0,2\pi/L\}$.
Here  the distribution unit is $L^{-3}$.
}
\label{distribution}
\end{figure}
It is interesting to see how these polarons or molecules look like from weak to strong interactions.
For this purpose, we need to plot the wave function in the real space.
Due to the reason that the visual dimensions are  restricted to up to three dimension,  therefore we can at most plot the wave function
of three fermions.  To further simplify the situation,   { we plot the probability distribution
for fixing  one of the fermions at $x_3=L/2$.}  From Eq.(\ref{total}),  we
have the distribution  function
\begin{eqnarray}
&&\rho(x_1,x_2)=
|f_{tot}(x_1,x_2,L/2)|^2\nonumber\\
&=&\frac{1}{G}\Big\{|f_{\downarrow_1}(x_1,x_2,L/2)|^2+|f_{\downarrow_2}(x_1,x_2,L/2)|^2\nonumber\\
&&+|f_{\downarrow_3}(x_1,x_2,L/2)|^2\Big\}\,.
\label{distributions}
\end{eqnarray}
{Here $G$ is the normalization factor. }

We focus on  the ground state. In the case of $c=0$,
this means that we have $\{k_1,k_2,k_3\}=\{0,0,2\pi/L\}$ and
\begin{eqnarray}
\rho(x_1,x_2)&={\frac{1} {L^3}} \Big\{&1+\frac{1}{3}\cos(\frac{2\pi}{L}x_1)+
\frac{1}{3}\cos(\frac{2\pi}{L} x_2)\nonumber\\
&&-\frac{1}{3}\cos[\frac{2\pi}{L} (x_1- x_2)]\Big\}\,.
\end{eqnarray}
The probability distributions are plotted for four different values of $c$.
in Fig. \ref{distribution}.  When $c=0$, the probability is the smallest one in the vicinity of  $x_1=x_2=x_3$ since the three fermions do not like to cluster together.
For the weak attraction, $c=-1$, the probability distribution
looks very similar to the case $c=0$. However, {as the attraction gets stronger,
the distribution changes dramatically. As shown in Fig. \ref{distribution},
for $c=-4$, there is a concentrated red area around the point  $x_1=x_2=x_3$, implying that
the two fermions form a  bound pair which tends to stay with the excess fermion under a strong attraction.} This is a clear indication of the crossover region from  a polaron to a molecule.
When the attraction is very strong, e.g., $c=-400$, the probability is significantly different
from zero only in a small area around $x_1=x_2=x_3$ and around the three lines,
$x_1=x_3, x_2=x_3, x_1=x_2$. This signals the formation of a molecule.
Note that the probability is exactly zero at $x_1=x_2=x_3$ due to the Fermi statistics.

\section{CORRELATIONS}
With the wave function in Eq.(\ref{total}), we can compute various correlation
functions, from which we can gain more insights into  the properties of polarons
and molecules.   We focus on the one-body correlation function~\cite{Chang2009,Leeuw2014}
and the density-density correlation function~\cite{Caux2006,Yu2009}.

\subsection{One-body correlation function}
The one-body correlation function is in fact the reduced one-body density matrix. It can be regarded as the probability of creating  a particle at the  position $x$ while annihilating a particle at the position  $x'$ at the same time.
For our system, there are two types of such a correlation function,
${\langle}a_{\uparrow}^{+}(x)a_{\downarrow}(x'){\rangle}$ and
${\langle}a_{\uparrow}^{+}(x)a_{\uparrow}(x'){\rangle}$. The former is clearly zero
since we cannot annihilate one spin-down fermion at $x'$ and
create one spin-up fermion at $x$.  For the latter, without loss of generality,
we set $x'=L/2$ and have
\begin{eqnarray}
&&{\langle}a_{\uparrow}^{+}(x)a_{\uparrow}(\frac L2){\rangle}={\frac{1} {G}}\int\cdots\int dx_{2}\cdots dx_{N}\Big\{ \nonumber\\
&&\sum_{j=2}^Nf_{\downarrow_j}^{*}(x,x_{2},\cdots,x_{N})f_{\downarrow_j}
(\frac L2,x_{2},\cdots,x_{N})\Big\}\,.
\label{onebody}
\end{eqnarray}
Here the wave function was  normalized.

\begin{figure}
\includegraphics[width=0.95\linewidth]{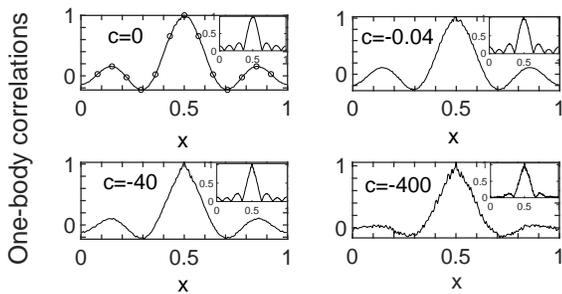}
\caption{ One-body correlations of the ground state at $c=0, -0.04, -40, -400$.
When $c=0$,  the momenta are $\{0, 0, \pm2\pi, \pm4\pi, \pm6\pi\}$. The black lines are the numerical results of the real part of the correlation function; the hollow circles denote the analytical results of $c=0$. The insets show the absolute values of the one-body correlations for $N=8$.  Here  the unit of correlations is $L^{-1}$}.
\label{one body correlation}
\end{figure}

When $c=0$, we can easily  calculate the correlation function
\begin{eqnarray}
{\langle}a_{\uparrow}^{+}(x)a_{\uparrow}(\frac L2){\rangle}
&=& {\frac{1} {L}} \Big\{\frac{(-1)^{\frac{N-2}{2}}}{N-1}\frac{\cos[\frac{(N-1)\pi x}{L}]}
{\cos(\frac{\pi x}{L})}\Big\}\nonumber\\
&\approx&{\frac{1} {L}} \Big\{\frac{(-1)^{\frac{N}{2}}}{N}\frac{\cos(k_F x)}{\cos(\frac{\pi x}{L})}\Big\}\,,
\label{frie}
\end{eqnarray}
where $k_F=N\pi/L$ is the Fermi wave-vector of the system.
When $c\neq 0$, we have to rely on  the numerical method, except the limit cases \cite{Nandani:2016}.
The  multi-dimensional integration in Eq.(\ref{onebody}) is done
with the Monte Carlo (MC) method and the results for $N=8$ are shown
Fig. \ref{one body correlation}.  In general, the one-body correlation function
is complex. For the cases studied here, the imaginary parts of the
correlation functions are small. Therefore, shown in Fig. \ref{one body correlation} are
the real parts of the one-body correlation functions; the absolute values
are shown in the insets.

The correlation functions for different values of interaction strength in Fig. \ref{one body correlation} show  peaks
around $x=L/2$ and then decay off over large distance, a common feature
of all correlation functions. Their decay tails are oscillatory. The oscillation
period is $2\pi/k_F$, as indicated in Eq.(\ref{frie}), and it changes little
as the attractive interaction gets stronger.   The absolute values of the correlation function show in the insets,
where the oscillation period is  $\pi/k_F$, which is the same as
the Friedel oscillations\cite{Mao2011}.  This is expected as the
cause of the oscillations in Fig. \ref{one body correlation}  and
also as the cause of the Friedel oscillations.

Another interesting feature in Fig. \ref{one body correlation} is that
the correlation decays faster when  the attractive interaction get stronger.
\subsection{Density-density correlation functions}
\begin{figure}
\includegraphics[width=0.95\linewidth]{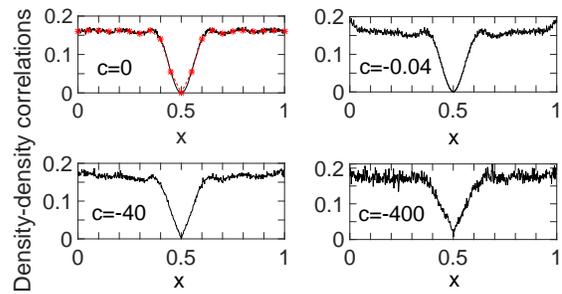}
\caption{(color online) Density-density correlations between up spins in the ground sate
for  $c=0, -0.04, -40, -400$.  When $c=0$, the momenta are
$\{0, 0, \pm2\pi, \pm4\pi, \pm6\pi\}$. The black lines are the numerical results;
the red line is the analytical result of $c=0$.
 $N=8$. The unit of length is $L$ { and the unit of correlations is $L^{-2}$}.}
\label{density-density(up-up) correlations}
\end{figure}

There are two types of density-density correlations, one between up spins
and the other between up and down spins. The density-density correlation between
up spins is
\begin{eqnarray}
&&{\langle}a_{\uparrow}^{+}(x)a_{\uparrow}(x)a_{\uparrow}^{+}(\frac L2)a_{\uparrow}(\frac L2){\rangle}= \int\cdots\int dx_{3}...dx_{N}\nonumber\\
&&{\frac{1} {G}}\Big\{\sum_{j=3}^Nf_{\downarrow_j}^{*}(x,\frac L2,x_3,\cdots,x_{N})f_{\downarrow_j}(x,\frac L2,x_3,\cdots,x_{N})\Big\}\,.\nonumber\\
\end{eqnarray}
This correlation function indicates  a probability to find a spin-up fermion at $x$
when  there is a spin-up fermion at $\frac{L}{2}$. When $c=0$,  this correlation
function is calculated in a straight forward way
\begin{eqnarray}
&&{\langle}a_{\uparrow}^{+}(x)a_{\uparrow}(x)a_{\uparrow}^{+}(\frac L2)a_{\uparrow}(\frac L2){\rangle}\nonumber\\
&=&{\frac{1} {L^{2}}} \Big\{\frac{N-1}{N-2}-\frac{\cos[\frac{(2N-3)\pi x}{L}]}{2(N-1)(N-2)\cos(\frac{\pi x}{L})} \nonumber\\
&&+\frac{\sin(\frac{\pi x}{L})\sin[\frac{(2N-3)\pi x}{L}]-1}{2(N-1)(N-2)\cos^{2}(\frac{\pi x}{L})}\Big\} \nonumber\\
&\approx&{\frac{1} {L^{2}}} \Big\{1+\frac{\cos(2k_F x+\frac{\pi x}{L})}{2N^2\cos(\frac{\pi x}{L})}  \nonumber\\
&&+\frac{\sin(\frac{\pi x}{L})\sin(2k_F x+\frac{\pi x}{L})-1}{2N^2
\cos^{2}(\frac{\pi x}{L})} \Big\}\,.
\end{eqnarray}
As shown in Fig.\ref{density-density(up-up) correlations},  this function is zero
at $x=L/2$ and approaches a constant when $x$ is far away from $L/2$.
The zero value of the density-density correlation at $x=L/2$ is due to the Pauli exclusion principle:
there can only be one fermion with up spin at $x=L/2$.  When $x$ is
far away from $L/2$,  the effect of the exclusion principle becomes weak
and the probability to find another up-spin fermion becomes a constant
as the system is in uniform.

\begin{figure}
\includegraphics[width=0.95\linewidth]{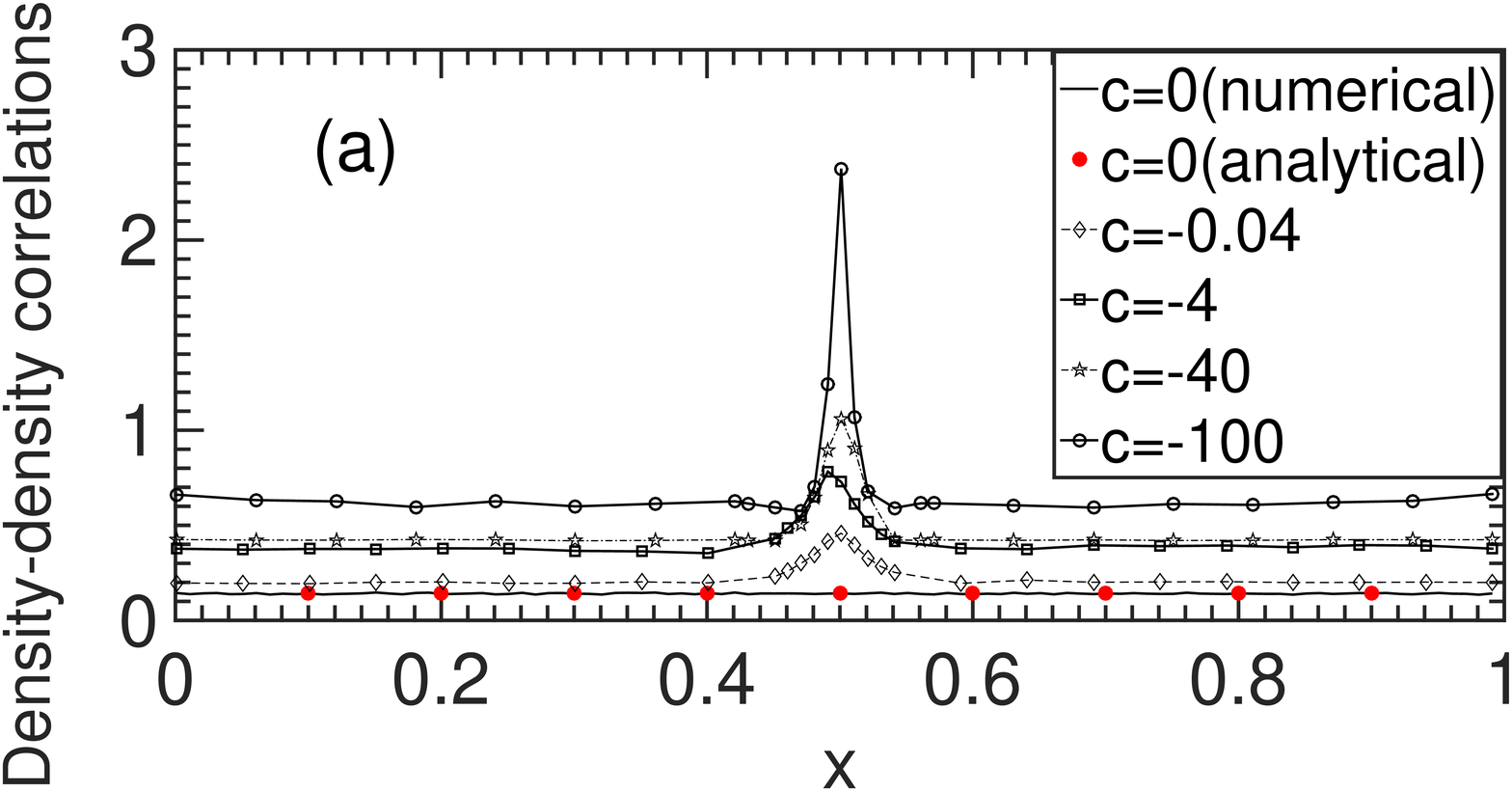}
\includegraphics[width=0.95\linewidth]{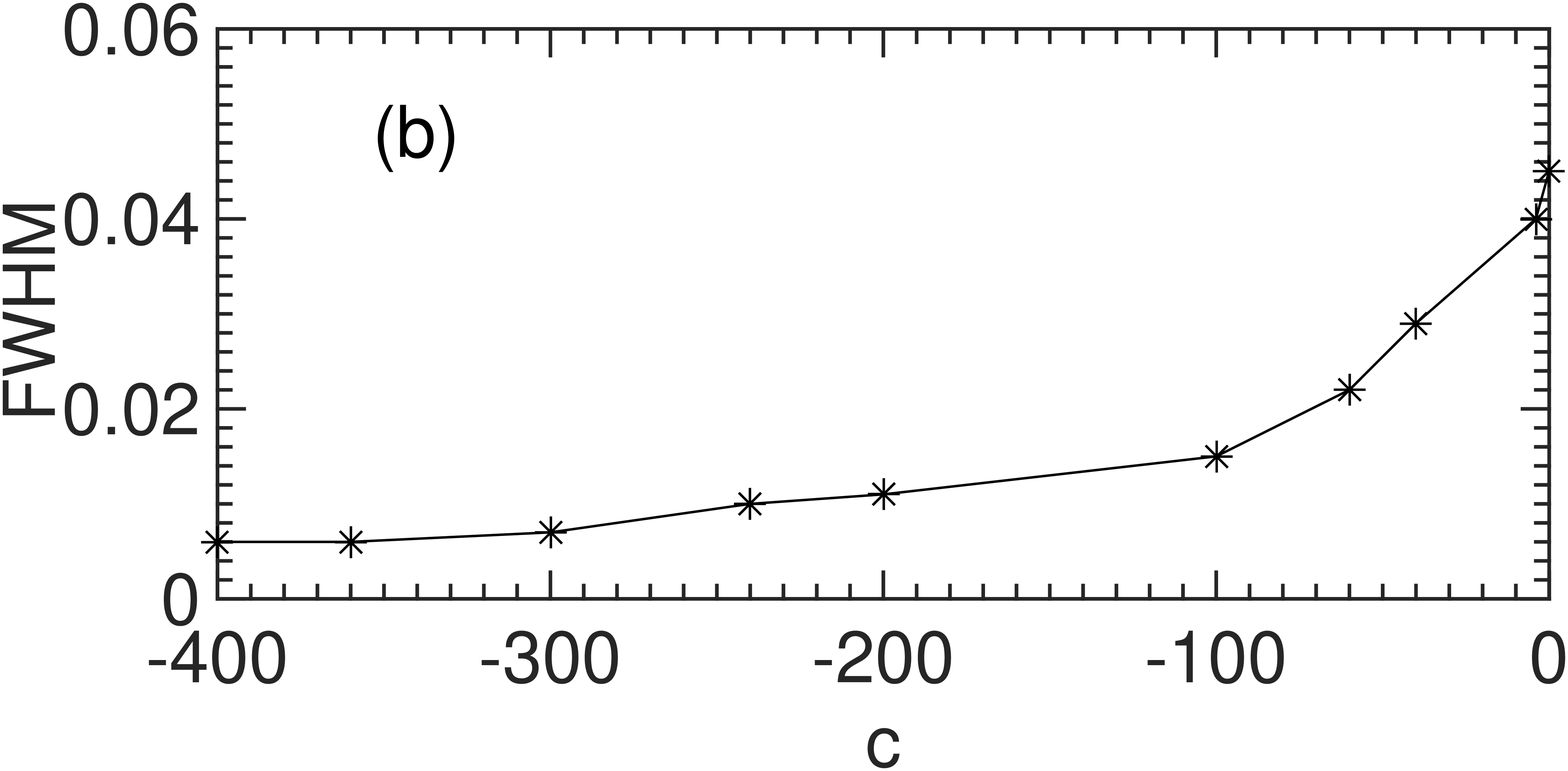}
\caption{(color online) (a) Density-density correlations between up spin and 
down spin in the ground sate
for $c=0, -0.04, -4, -40, -100$. For clarity, the curves for $c\neq 0$ have shifted upwards by
$0.1, 0.2, 0.3, 0.4$, respectively.  When $c=0$, the momenta are
$\{0, 0, \pm2\pi, \pm4\pi, \pm6\pi\}$. The different symbol lines are the numerical 
results for different interaction;
the red circles are the analytical result for $c=0$.
(b) The  full width at half maximum (FWHM) as a function of the interaction strength $c$.
$N=8$. The unit of length is $L$, the unit of correlations is $L^{-2}$ 
and the unit of $c$ is $L^{-1}$.}
\label{density-density(up-down) correlations1}
\end{figure}

When $c\neq 0$, the density-density correlations are computed numerically and
the results are plotted in Fig.\ref{density-density(up-up) correlations} for $N=8$.
We observe in  this figure that the interaction does not change the overall feature
of the correlation function. It appears that the amplitude of the oscillatory
tail is suppressed by the strong  interaction.
However, due to the limitation of the accuracy of
the numerical results,  it is hard to quantify this suppression.

The density-density correlations between up spin and down spin is
\begin{eqnarray}
&&{\langle}a_{\uparrow}^{+}(x)a_{\uparrow}(x)a_{\downarrow}^{+}(\frac L2)
a_{\downarrow}(\frac L2){\rangle}\nonumber\\
&&={\frac{1} {G}}\sum_{j=2}^{N-1}\int\cdots\int dx_2\cdots dx_{j-1}dx_{j+1}\cdots dx_{N}\nonumber\\
&&\Big\{f_{\downarrow_N}^{*}(x,x_2,\cdots,x_{j-1},\frac L2,x_{j+1},\cdots,x_N)\times\nonumber\\
&&f_{\downarrow_N}(x,x_2,\cdots,x_{j-1},\frac L2,x_{j+1},\cdots,x_N)\Big\}\,,\nonumber\\
\end{eqnarray}
which gives us the probability to find a spin-up fermion at $x$ when
there is a spin-down fermion at  $\frac{L}{2}$.   When $c=0$,
this density-density correlations is a constant and equal to $1/[(N-1)L^2]$.
This reflects the fact that when there is no interaction the presence of a down-spin
fermion does not affect the density of up-spin fermions. When $c\neq 0$, the correlation
functions are computed numerically and the results are shown in
Fig.\ref{density-density(up-down) correlations1} (a) for $N=8$. From the figure,
we see that a peak immediately emerges when the attractive interaction is turned on.
This is resulted from the formation of polaron. As the interaction gets stronger and  stronger,
the peak becomes narrower and narrower. This signature indicates  a crossover from 
the polaron-like behaviour to the tightly bound  molecule.
Furthermore,  the full widths at half maximum (FWHM) as a function of interaction strength
is plotted in Fig.\ref{density-density(up-down) correlations1} (b).

The density-density correlation function charaterizes the  interference  of two particles. The measurement of many-body correlations was carried out for bosons via a single atom detector, where the many-body  Wick's theorem provides a significant theoretical input \cite{Dall:2013}. It is highly desirable to adapt this experimental technique to measure many-body correlations for interacting fermions. On the other hand, in the cold atoms experiments, one can overlap a TOF image with its copy that is shifted by $x$.
Integrating over the overlapped region, then  one may measure the density-density correlation of $x$.
In fact,  in current  experiments with ultracold atoms, the ratio frequency spectroscopy  of the ultracold atoms is often used to demonstrate the quasiparticle behaviour of the impurity problems \cite{Schirotzek2009,Navon2010,Kohstall:2012,Jorgensen:2016,Hu-M:2016}. The shifts and widths of the spectra are conveniently used to  read off the average binding energy and lifetime of polaron. To this end, the spectral function $A(k,\omega)$, which gives the probability to find the state with a frequency $\omega$ and momentum $\hbar k$, is a central importance in the problems of this kind. Once we turn on the interaction, the spectral function $A(k,\omega)$ thus contains Lorentzian peaks. However, the spectral function is related to the single-particle Green's function  (retarted Green's function), i.e., $A(k,\omega)=\frac{-1}{\pi}{\rm ImG}_{\rm ret}(k,\omega)$,   is cumbersome for computing   in general. We will consider this study  elsewhere.

\section{Conclusion}

We have studied the formation of polaron and molecule when one spin-down fermion
is placed in a sea of free up-spin fermions.  It shows that as  the attractive interaction between
up-spin and down-spin fermions increases, the spin-down fermions is dressed up by the surrounding ones  from the Fermi sea to form a  polaron. Whereas, for an strong attraction the single spin-down fermion tightly bounds one spin-up fermion to form a  molecule. We have obtained analytically the effective masses, binding energies  and kinetic energies of the polaron and molecule. We have also numerically calculated these key properties for a whole interacting regime. The numerical results confirm the novel nature of the crossover from polaron to molecule in the 1D impurity problem.  For the further study of this problem, we have numerically calculated  the probability distribution function and the
one-body and density-density correlation functions from which the nature of the polaron and molecule in 1D is demonstrated. Our results provide a precise understanding of such typical collective many-body phenomenon caused by quantum impurity. Our method can be directly apply to the impurity problems in different mediums with integrability.

\acknowledgements
We acknowledge helpful discussion with Yuzhu Jiang.
This work is supported by the NBRP of China (2013CB921903,2012CB921300) and
the NSF of China (11274024,11334001,11429402,11374331,11534014).

\begin{appendix}

\section{The four-body wave function}
The wave function for the four-fermion case is
\begin{eqnarray}
&&f_{\downarrow_4}(x_{1},x_{2},x_{3},x_{4})\nonumber\\
&=&\left|\begin{array}{ccc}
(k_{2}-\tilde{\lambda})e^{ik_{2}x_{1}} & (k_{3}-\tilde{\lambda})e^{ik_{3}x_{1}} & (k_{4}-\tilde{\lambda})e^{ik_{4}x_{1}}\\
(k_{2}-\tilde{\lambda})e^{ik_{2}x_{2}} & (k_{3}-\tilde{\lambda})e^{ik_{3}x_{2}} & (k_{4}-\tilde{\lambda})e^{ik_{4}x_{2}}\\
(k_{2}-\tilde{\lambda})e^{ik_{2}x_{3}} & (k_{3}-\tilde{\lambda})e^{ik_{3}x_{3}} & (k_{4}-\tilde{\lambda})e^{ik_{4}x_{3}}\end{array}\right|e^{ik_{1}x_{4}}\nonumber\\
&+&\left|\begin{array}{ccc}
(k_{3}-\tilde{\lambda})e^{ik_{3}x_{1}} & (k_{4}-\tilde{\lambda})e^{ik_{4}x_{1}} & (k_{1}-\tilde{\lambda})e^{ik_{1}x_{1}}\\
(k_{3}-\tilde{\lambda})e^{ik_{3}x_{2}} & (k_{4}-\tilde{\lambda})e^{ik_{4}x_{2}} & (k_{1}-\tilde{\lambda})e^{ik_{1}x_{2}}\\
(k_{3}-\tilde{\lambda})e^{ik_{3}x_{3}} & (k_{4}-\tilde{\lambda})e^{ik_{4}x_{3}} & (k_{1}-\tilde{\lambda})e^{ik_{1}x_{3}}\end{array}\right|e^{ik_{2}x_{4}}\nonumber\\
&+&\left|\begin{array}{ccc}
(k_{4}-\tilde{\lambda})e^{ik_{4}x_{1}} & (k_{1}-\tilde{\lambda})e^{ik_{1}x_{1}} & (k_{2}-\tilde{\lambda})e^{ik_{2}x_{1}}\\
(k_{4}-\tilde{\lambda})e^{ik_{4}x_{2}} & (k_{1}-\tilde{\lambda})e^{ik_{1}x_{2}} & (k_{2}-\tilde{\lambda})e^{ik_{2}x_{2}}\\
(k_{4}-\tilde{\lambda})e^{ik_{4}x_{3}} & (k_{1}-\tilde{\lambda})e^{ik_{1}x_{3}} & (k_{2}-\tilde{\lambda})e^{ik_{2}x_{3}}\end{array}\right|e^{ik_{3}x_{4}}\nonumber\\
&+&\left|\begin{array}{ccc}
(k_{1}-\tilde{\lambda})e^{ik_{1}x_{1}} & (k_{2}-\tilde{\lambda})e^{ik_{2}x_{1}} & (k_{3}-\tilde{\lambda})e^{ik_{3}x_{1}}\\
(k_{1}-\tilde{\lambda})e^{ik_{1}x_{2}} & (k_{2}-\tilde{\lambda})e^{ik_{2}x_{2}} & (k_{3}-\tilde{\lambda})e^{ik_{3}x_{2}}\\
(k_{1}-\tilde{\lambda})e^{ik_{1}x_{3}} & (k_{2}-\tilde{\lambda})e^{ik_{2}x_{3}} & (k_{3}-\tilde{\lambda})e^{ik_{3}x_{3}}\end{array}\right|e^{ik_{4}x_{4}}\nonumber\\
\end{eqnarray}
where $\tilde{\lambda}=\lambda-ic'\mathrm{sign}(x_{4}-x_{i}),$ ($i$ is the indice of $k$ in one term).


\section{Analytical solutions of the Bethe ansatz equations}
In this Appendix, we offer detailed derivation for the solutions of the Bethe
ansatz (BA) equations Eqs.(\ref{ba01},\ref{ba02}) up to the second order in both the
weak and strong interaction limits. To simplify the notations, we  make the following
change of variables:
\begin{eqnarray}
\tilde{k}_j=k_jL\,, ~~~\tilde{c}'=c'L\,,  ~~~\tilde{\lambda}=\lambda L\,.
 \end{eqnarray}
Eqs.(\ref{ba01},\ref{ba02})  then become
\begin{eqnarray}
&&\frac{\tilde{k}_j-\tilde{\lambda}+i\tilde{c}'}{k_j-\lambda-i\tilde{c'}}=\exp(i\tilde{k}_j)\,,\\
&&\prod_{j=1}^N\frac{\tilde{k}_j-\tilde{\lambda}+i\tilde{c}'}{\tilde{k}_j-\tilde{\lambda}-i\tilde{c}'}=1\,,
\end{eqnarray}
Without causing confusion and for convenience, we drop tilde first and recover it at the end.
That is that we now use
\begin{eqnarray}
&&\frac{k_j-\lambda+ic'}{k_j-\lambda-ic'}=\exp(ik_j)\label{BA3}\,,\\
&&\prod_{j=1}^N\frac{k_j-\lambda+ic'}{k_j-\lambda-ic'}=1\,.
\label{BA4}
\end{eqnarray}
There are infinite solutions. We focus on the solutions near the ground state.

\subsection{Weak interaction limit}
In the ground state of the non-interacting case $c=0$,  the up-spin fermions occupy
the momentum states $\{0,\pm 1, \pm 2,\cdots,\pm (N-2)/2\}2\pi$ ($N$ is assumed to be even in the main text)
and the sole down-spin fermion has zero momentum. The small excitation state correspond to that
the down-spin takes up non-zero momentum. In all these states, there is only one momentum state  that
is occupied by both up-spin and down-spin fermions. If this momentum state is $k_p=2n_p\pi$, the total
momentum $q$ of the system is given by this momentum and we have $q=2n_p\pi$.
Our focus is on these states.

\def\be{\begin{equation}}
\def\ee{\end{equation}}
\def\ba{\begin{eqnarray}}
\def\ea{\end{eqnarray}}

We now turn on a small attractive interaction. The common momentum shared by the up
and down-spin is split into two $k_p^{\pm}=p\pm i\beta$ while all the other momenta
are shifted as $k_j=2n_j\pi+\delta k_j$. Our aim is to compute $p$, $\beta$, and $\delta k_j$.
Note that despite of the change of each momentum $k_j$ by the interaction, the total momentum
$q$ of the system remains unchanged. We still have $q=2n_p\pi$. This is because
that the interaction is between fermions and is incapable of changing the total momentum.

For $k_j=2n_j\pi+\delta k_j$, we have from Eq.(\ref{BA3})
\ba
\exp(ip-\beta)&=&\frac{p-\lambda+i(\beta+c')}{p-\lambda+i(\beta-c')}\,,\\
\exp(ip+\beta)&=&\frac{p-\lambda-i(\beta-c')}{p-\lambda-i(\beta+c')}\,,
\ea
which lead to
\begin{eqnarray}
\tan p&=&\frac{2c'(p-\lambda)}{(p-\lambda)^2+(\lambda^2-c'^2)}\label{BA5}\,,\\
(p-\lambda)^2&=&-\beta^2-c'^2+c'\frac{2\lambda (e^{-2\beta}+1)}{e^{-2\beta}-1}\,.
\label{BA6}
\end{eqnarray}
So, when $c\rightarrow 0$, $(p-\lambda)\rightarrow 0$. Since $p=2n_p\pi$ at $c=0$,
we know that $\lambda\rightarrow 2n_p\pi$ as $c\rightarrow 0$.  This means that
$k_j-\lambda\sim 2\pi$.  Knowing this fact of the weak coupling limit,
we have from Eq.(\ref{BA3}) to the first order of $c$
\begin{eqnarray}
\frac{1+\frac{ic'}{k_j-\lambda}}{1-\frac{ic'}{k_j-\lambda}}&=&\exp(ik_j)\,,
\label{BA10}\\
1+\frac{ic}{2(n_j-n_p)\pi}&\approx&1+i(k_j-2n_j\pi)\,,
\end{eqnarray}
We then have
\be
k_j\approx 2n_j\pi-\frac{|c|}{2(n_j-n_p)\pi}\,.
\ee
As the total momentum $q=2p+\sum_jk_j=2n_p\pi$,  we have
\begin{eqnarray}
p\approx2n_p\pi+\frac{1}{2}\sum_{j=1}^{N-2}\frac{|c|}{2(n_j-n_p)\pi}\,.
\end{eqnarray}
Note that in the above summation and any following summation involving $n_j-n_p$ we always
assume that $n_j\neq n_p$.

After cancelling $\lambda$ from Eqs.(\ref{BA5},\ref{BA6})  we have
\ba
&&-\beta^2-c'^2+ c' \frac{2\beta}{e^{-2\beta}-1}(e^{-2\beta}+1)\nonumber\\
&=&\tan^2 p \Big(\frac{e^{-2\beta}+1}{e^{-2\beta}-1}\beta-c' \Big)^2\,.
\label{BA14}
\ea
Since $\tan^2 p=\tan^2 (p-2n_p\pi)\sim c^2$,  we have to have $\beta\approx \sqrt{|c|}$.
So, the system energy up to the first order of $c$ is
\begin{eqnarray}
E &=&\frac{\hbar^{2}}{2m}\Big( -2\beta^{2}+2p^{2}+\sum_{j=1}^{N-2}k_{j}^{2}\Big)\nonumber\\
&\approx&-\frac{\hbar^{2}}{2m}\frac{2(N-1)|c|}{L}+
\frac{\hbar^{2}q^{2}}{2m}+E_{0},\label{polaron}
\label{polaronenergy}
\end{eqnarray}

We have now computed $p$, $k_j$ and $\beta$ to the lowest order.
We next try to compute $p$, $k_j$ and $\beta$ to the next order. We write
\begin{eqnarray}
p&=&2n_{p}\pi+p^{(1)}+p^{(2)}+\cdots\\
k_{j}&=&2n_{j}\pi+k_{j}^{(1)}+k_{j}^{(2)}+\cdots\\
\beta&=&\sqrt{|c|}+\beta^{(2)}\,,
\end{eqnarray}
where $p^{(1)}$ and $k_{j}^{(1)}$ are already computed in the above.
We have from Eq.(\ref{BA10})
\begin{eqnarray}
\frac{1+\frac{ic'}{2n_j\pi+k_j^{(1)}-2n_p\pi-2p^{(1)}}}{1-\frac{ic'}{2n_j\pi+k_j^{(1)}
-2n_p\pi-2p^{(1)}}}\approx\exp(ik_j)\\
\frac{2ic'}{2n_j\pi-2n_p\pi}- \frac{2i(k_j^{(1)}-2p^{(1)})c'}{(2n_j\pi-2n_p\pi)^2}\nonumber\\
+2\Big(\frac{ic'}{2n_j\pi-2n_p\pi}\Big)^2\approx i(k_j^{(1)}+k_{j}^{(2)})-\frac{(k_j^{(1)})^2}{2}\,.
\label{zhankai3}
\end{eqnarray}
This leads to
\begin{eqnarray}
k_j^{(2)}&=&-\frac{c^{2}}{8(n_{j}-n_{p})^{2}\pi^3}\{\frac{1}{n_{j}-n_{p}}\nonumber\\
&+&\sum_{i=1}^{N-2}\frac{1}{n_{i}-n_{p}}\}\,.
\ea
We again use that $q=2p+\sum_jk_j=2n_p\pi$ to find
\ba
p^{(2)}&=&\frac{c^{2}}{16\pi^{3}}\{\sum_{j=1}^{N-2}\frac{1}{(n_{j}-n_{p})^{3}}\nonumber\\
&+&\sum_{j=1}^{N-2}\sum_{i=1}^{N-2}\frac{1}{(n_{j}-n_{p})^{2}(n_{i}-n_{p})}\}\,.
\end{eqnarray}
Through the Taylor expansion, we have from Eq.(\ref{BA14})
\begin{eqnarray}
&&-\beta^2-c'^2-2c'\frac{1-\beta+\beta^2}{1-\beta+2\beta^2/3}\approx(p^{(1)})^2\,,\\
&&-\beta^2-c'^2+|c|(1+\frac{\beta^2}{3})\approx(p^{(1)})^2\,.
\end{eqnarray}
From this we obtain
\be
\beta^{(2)}\approx\frac{|c|^{3/2}}{24}-\frac{(p^{(1)})^{2}}{2\sqrt{|c|}}.
\ee
With the above results, we have the second-order energy
\begin{eqnarray}
E^{(2)}=-\frac{\hbar^{2}}{2m}\Big\{\frac{c^{2}}{6}+
\frac{c^{2}}{2\pi^{2}}\sum_{j=1}^{N-2}\frac{1}{(n_{j}-n_{p})^{2}}\Big\}.
\end{eqnarray}
Note that in the above summation we have $n_j\neq n_p$ and the units have
been restored.  It is not clear how to
extract a term which is proportional to $n_p^2$ and obtain the second order correction
to the effective mass.

\subsection{Strong interaction limit}
We now consider the strong coupling limit, i.e., $|c|\gg1$, where we compute
everything up to the first order of $1/|c|$.  In this limit we have $\beta\gg 1$
and thus $e^{-2\beta}\ll 1$.  From Eq.(\ref{BA6}), we have
\ba
(p-\lambda)^2&\approx&-\beta^2-c'^2-2\beta c' (1+e^{-2\beta})^2\nonumber\\
&\approx&-(\beta+c')^2-4\beta c' e^{-2\beta}\nonumber\\
&\approx&-(\beta+c')^2\,.
\ea
This gives us
\be
p\approx \lambda\,,~~~~~\beta\approx -c'=|c|/2\,.
\ee

In the limit of $|c|\rightarrow \infty$,  if the system is stable near the ground state,
$k_j$'s and $p$ must be finite.  With Eq.(\ref{BA3})  this implies that
$\exp(ik_j)\rightarrow -1$ and $k_j\rightarrow n_j\pi$ with $n_j$ being an odd integer.
Combining Eq.(\ref{BA3}) and Eq.(\ref{BA4}) we have $e^{-2ip}=e^{i\sum_jk_j}$.
This means that we have $e^{2ip}\rightarrow (-1)^{N-2}$ in the limit of $|c|\rightarrow \infty$.
As $N$ is even (which is assumed in this work),  we have $p\rightarrow n_p\pi$
with $n_p$ being an arbitrary integer.

With the above results we have
\begin{eqnarray}
\exp(ik_j)&\approx&\frac{k_j-p+ic'}{k_j-p-ic'}\\
1+i\Big[k_j-n_j\pi\Big]&\approx&\frac{-i(n_j-n_p)\pi/c'+1}{i(n_j-n_p)\pi/c'+1}\\
k_{j}&\approx& n_{j}\pi+\frac{4\pi(n_j-n_p)}{|c|}\,.
\end{eqnarray}
with $n_{j}$ being odd integers. As $q=\sum_j k_j+2p=2n_p\pi$, we have
\be
p\approx
n_p\pi-\sum_{j=1}^{N-2}\frac{2n_j\pi}{|c|}+\frac{2(N-2)n_p\pi}{|c|}\,.
\ee
From all the above results, it is clear that the ground state of the system
in the limit of $|c|\rightarrow \infty$ corresponds to that $n_j$'s take
values of $\pm1,\pm3, \pm5,\cdots, \pm (N-3)$ while $n_p=0$. $n_p\neq =0$
corresponds to excited states.

So the system's energy up to the order $1/|c|$ is
\begin{eqnarray}
E&=&\frac{\hbar^{2}}{2m}\Big(-2\beta^{2}+2p^{2}+\sum_{j=1}^{N-2}k_{j}^{2}\Big)\nonumber\\
&=&-\frac{\hbar^{2}}{2m}\Big[\frac{c^{2}}{2}+(1+\frac{8}{|cL|})\frac{(2N-1)(N-2)\pi^{2}}{L^2}\Big]\nonumber\\
&&+\Big[\frac{1}{2}+\frac{2(N-2)}{|cL|}\Big]\frac{\hbar^{2}q^{2}}{2m}+(1+\frac{8}{|cL|})E_{0}\,.
\label{moleculeenergy}
\end{eqnarray}
The effective mass can be extracted from the kinetic part of the energy and it is
\begin{equation}
m^{*}\approx2m\Big[1-\frac{4(N-2)}{L|c|}\Big]\,.
\end{equation}
\end{appendix}
\end{document}